\documentclass[aps,prl,reprint, amsmath, amssymb,superscriptaddress]{revtex4-1}

\usepackage{dsfont}
\usepackage{graphicx} 
\usepackage{graphics}

\usepackage{mathtools}
\usepackage{dcolumn}    
\usepackage{bm} 
\usepackage{graphicx}
\usepackage{amsmath}    
\usepackage{latexsym}
\usepackage{amsfonts}   
\usepackage{amssymb}
\usepackage{array}      
\usepackage{epsfig}
\usepackage{txfonts}
\usepackage{xcolor}
\usepackage[normalem]{ulem}
\usepackage[colorlinks=true,linkcolor=blue,urlcolor=blue,citecolor=blue,pdfusetitle]{hyperref}
\usepackage{hyperref}
\usepackage{framed}

\newcommand{\Tr}{\mbox{Tr}}

\newcommand{\norm}[1]{\left\lVert#1\right\rVert}

\newcommand{\ket}[1]{\left|#1\right\rangle}
\newcommand{\bra}[1]{\langle#1|}
\newcommand{\braket}[2]{\langle#1|#2\rangle}
\newcommand{\ketbra}[2]{|#1\rangle\langle#2|}

\newcommand{\avg}[1]{\langle#1\rangle}

\newcommand{\Dket}[1]{\left| #1 \right)}
\newcommand{\Dbra}[1]{\left( #1 \right|}
\newcommand{\Dbraket}[2]{\left( #1 | #2 \right)}
\newcommand{\Dketbra}[2]{\left| #1 \right)\left( #2 \right|}

\begin{document}

\title{Universally Robust Quantum Control}
\author{Pablo M. Poggi}
\email{pablo.poggi@strath.ac.uk }
\affiliation{Department of Physics, SUPA and University of Strathclyde, Glasgow G4 0NG, United Kingdom }
\affiliation{Center for Quantum Information and Control, Department of Physics and Astronomy, University of New Mexico, Albuquerque, New Mexico 87131, USA}
\author{Gabriele De Chiara}
\affiliation{Centre for Quantum Materials and Technology, School of Mathematics and Physics, Queen’s University Belfast, Belfast BT7 1NN, United Kingdom}
\author{Steve Campbell}
 \address{School of Physics, University College Dublin, Belfield Dublin 4, Ireland}
 \affiliation{Centre for Quantum Engineering, Science, and Technology,
University College Dublin, Belfield, Dublin 4, Ireland}
\affiliation{Dahlem Center for Complex Quantum Systems, Freie Universit\"at Berlin, Arnimallee 14, 14195 Berlin, Germany}
 \author{Anthony Kiely}
 \address{School of Physics, University College Dublin, Belfield Dublin 4, Ireland}
 \affiliation{Centre for Quantum Engineering, Science, and Technology,
University College Dublin, Belfield, Dublin 4, Ireland}

\begin{abstract}
We study the robustness of the evolution of a quantum system against small uncontrolled variations in parameters in the Hamiltonian. We show that the fidelity susceptibility, which quantifies the perturbative error to leading order, can be expressed in superoperator form and use this to derive control pulses which are robust to any class of systematic unknown errors. The proposed optimal control protocol is equivalent to searching for a sequence of unitaries that mimics the first-order moments of the Haar distribution, i.e. it constitutes a 1-design. We highlight the power of our results for error-resistant single- and two-qubit gates.
\end{abstract}

\maketitle

{\bf \emph{Introduction.--}}
Tremendous advances in the ability to manipulate states of light and matter are ushering in the new generation of quantum-enhanced devices. As recently remarked~\cite{OCreview2}, it is precisely the ability to develop schemes to control a system that endows scientific knowledge with the potential to revolutionise technological landscapes~\cite{DeutschPRXQ, Preskill2018}. However, while exquisite levels of control are now routinely applied in a variety of platforms~\cite{larrouy2020,lysne2020,werninghaus2021}, there will always be systematic errors due to imperfect fabrication and incomplete knowledge of the parameters, either in relation to the model itself or the ambient conditions under which it is operating. Thus, several strategies to explicitly mitigate such errors have been devised, e.g. shortcuts to adiabaticity \cite{STAreview,ruschhaupt2012,kiely2014}, numerical optimization~\cite{OCreview2,anderson2015,Coopmans2022}, geometric space curves~\cite{Dong2021,Buterakos2021,Barnes2022}, composite pulses~\cite{CompositePulses1,CompositePulses2}, and dynamical decoupling~\cite{lidar2014review}. 

When these systematic errors are important, typically the control problem is cast in such a way that two, sometimes implicit, assumptions are made regarding the source of the error: (i) that it arises from a weak perturbation, and (ii) that its mathematical structure is exactly known. While the former is a reasonable working condition to assume (if it were not then the fundamental description of the system  would need to be adjusted), the latter is arguably less well justified. Indeed, concerted effort is currently invested in identifying the correct physical description of noisy intermediate-scale quantum devices, e.g. determining the most relevant noise sources that they are subject to in order to enhance their efficacy~\cite{Rudinger2021}. Ultimately, there will always be some level of uncertainty in our knowledge of the precise structure of the noise and therefore it is highly desirable to develop a framework that allows to coherently manipulate quantum systems even in the presence of an unknown (even possibly unknowable) source of error. 

In this work, we develop such a framework which accounts for this uncertainty, termed universally robust control (URC). It provides a straightforward cost function to be minimised to ensure generic robustness in quantum control problems. It can also be easily restricted to specific classes of errors, to account for a limited but useful knowledge of the error type. 

{\bf \emph{Fidelity in the presence of systematic error.--}}Consider the full system Hamiltonian $H_\lambda(t)\!=\!H_0(t)+\lambda V$ where $H_0(t)$ is the error-free control Hamiltonian, $V$ is the error operator acting with unknown strength, $\lambda$. We assume a pure initial state, $\sigma$, with no $\lambda$ dependence.

The time evolution operator of $H_\lambda(t)$ is given by $U_\lambda(t,0)$, which leads to the $\lambda$-dependent state $\rho_\lambda\!=\!U_\lambda(t_f,0) \sigma U^\dagger_\lambda(t_f,0)$ at the final time $t \!=\! t_f$. The fidelity between the perturbed and ideal evolution is $F(\lambda)\!=\!\Tr(\rho_\lambda \rho_0)$, which can be expanded for small $\lambda$ as
\begin{eqnarray}
F(\lambda) \approx F(0)+F'(0)\lambda + \frac{1}{2} F''(0) \lambda^2.
\label{taylor}
\end{eqnarray}
By definition $F(0)=1$ and from this follows $F'(0)=0$ \cite{SupMat}.

The second derivative can be calculated by noting that, for pure states, $\partial_\lambda^2 \rho_\lambda \!=\!2 \left(\partial_\lambda \rho_\lambda\right)^2+\rho_\lambda \left(\partial_\lambda^2 \rho_\lambda\right)+ \left(\partial_\lambda^2 \rho_\lambda\right) \rho_\lambda$. Multiplying by $\rho_0$ and evaluating the trace at $\lambda\!=\!0$ we get
\begin{eqnarray}
F''(0)=  -2\: \chi_S (\rho_\lambda), \label{d2}
\end{eqnarray}
where $\chi_S (\rho_\lambda)= \Tr \left\{ \rho_0 \left( \partial_\lambda \rho_\lambda\right)^2 \big\lvert_{\lambda=0} \right\}$ is the fidelity susceptibility \cite{Lu2008,Giorda2010,liu2020}, which quantifies how sensitive the evolution is with respect to small perturbations, i.e. $F(\lambda)\!\simeq\! 1-\chi_S(\rho_\lambda)\lambda^2$. It is clear that $\chi_S(\rho_\lambda)$ is simply the quantum Fisher information (QFI)  associated to the family of states $\{\rho_\lambda\}$ \cite{SupMat}. The QFI quantifies how much information about $\lambda$ is encoded in the evolution of the state, thus minimizing the QFI at $\lambda=0$ is equivalent to increasing the robustness of a control protocol.

 Evaluating explicitly the QFI we find~\cite{SupMat} 
\begin{equation}
    \chi_S(\rho_\lambda)=\frac{t_f^2}{\hbar^2}\left(\Delta \overline{V}_0
    \right)^2,
    \label{eq:state_qfi}
\end{equation}
where
\begin{eqnarray}
    \overline{V}_0 = \frac{1}{t_f}\int_0^{t_f} ds\:U_0^\dagger(s,0) V U_0(s,0),
    \label{eq:V0_def}
\end{eqnarray}
is the time average of $V$ in the interaction picture with respect to the unperturbed evolution and the variance is taken with respect to the initial state, $\left(\Delta \overline{V}_0\right)^2 = \Tr[\sigma \overline{V}_0^2]-\Tr[\sigma \overline{V}_0]^2$. 

A similar result can be derived for the case of the evolution of unitaries (instead of states). By defining the corresponding fidelity as $F_U(\lambda)\! = \! \frac{1}{d^2} \left\vert \mathrm{Tr}\left(U_0^\dagger U_\lambda\right)\right\vert^2$, we obtain that $F_U(\lambda)\simeq 1-\chi_U(U_\lambda)\lambda^2$ \cite{SupMat}. The susceptibility is
\begin{equation}
    \chi_U(U_\lambda) = \frac{t_f^2}{\hbar^2 d }||\overline{V}_0||^2,
    \label{eq:fide_sus_u}
\end{equation}
where $||\cdot||$ is the norm associated with the Hilbert-Schmidt inner product $\Dbraket{A}{B}=\Tr(A^\dagger B)$ and $d$ is the Hilbert space dimension. Robust control protocols then correspond to finding a $H_0(t)$ such that $\rho_0\!=\!\rho_{\mathrm{target}}$ or $U_0(t_f,0)=U_{\mathrm{target}}$ while concurrently minimizing $\chi_S$ for a known perturbation model $V$ \cite{ball2021,koswara2021,kosut2022}. We now demonstrate that such robust control can be achieved even \textit{without} knowledge of $V$.

{\bf \emph{Universally robust control.--}}Our construction is based on a superoperator picture where the operator 
\begin{equation}
    \mathcal{M}_{0}[V]\equiv \overline{V}_{0} ,
    \label{eq:superop_M}
\end{equation}
can be seen as the action of a (linear) superoperator $\mathcal{M}_{0}$ acting on $V$ and we assume that $\Tr V=0$~\footnote{If this were not the case we can always write $V=V'-(\Tr{V'}/d) \mathbb{I} $, where $\Tr{V}=0$. The term proportional to identity will not contribute to the dynamics besides an overall global phase which will not be relevant for control.}. To construct it more explicitly, we go to a doubled Hilbert space. If our original Hilbert space $\mathcal{H}$ is spanned by the orthonormal basis $\{\ket{i}\}$ where $i=1,\ldots,d$, we take
\begin{equation}
    A = \sum\limits_{ij} A_{ij} \ketbra{i}{j}\rightarrow \Dket{A} =\sum\limits_{ij} A_{ij} \ket{i}\otimes \ket{j},
\end{equation}
where $\Dket{A}$ lives in $\mathcal{H}\otimes\mathcal{H}$~\cite{SupMat}. From Eq. (\ref{eq:superop_M}) we define 
\begin{equation}
    M_{0} = \frac{1}{t_f}\int_0^{t_f} ds \left[U_0(s,0) \otimes U_0(s,0)^* \right]^\dagger.
    \label{eq:bare_suoperop}
\end{equation}
\noindent such that $\Dket{\overline{V}_{0}}=M_{0} \Dket{V}$.
The fidelity susceptibility of Eq. (\ref{eq:fide_sus_u}) can be expressed in terms of the superoperator $M_0$ as
\begin{equation}
    ||\overline{V}_{0}||^2 =\Dbra{V} M_{0}^\dagger M_{0} \Dket{V}.
    \label{eq:normV_superop}
\end{equation}

By virtue of Eq. (\ref{eq:fide_sus_u}) we can increase the robustness of a unitary control protocol irrespective of  $V$ by choosing $H_0(t)$ to minimize the operator norm of $M_{0}$. Intuitively, this is because $||M_0 \Dket{V}|| \leq  ||M_0|| \cdot || \Dket{V}||$. This also holds for state control, c.f. Eq. (\ref{eq:state_qfi}), because  $\Delta \overline{V}_0$ is upper bounded by $||M_0||$ ~\cite{SupMat}.  
 
The trace of any operator $V$ is unitarily invariant. For the identity operator $\mathbb{I}$, $M_0\Dket{\mathbb{I}}\!=\!\Dket{\mathbb{I}}$ so the norm of $M_0$ cannot be arbitrarily reduced. To sidestep this issue, we restrict to the set of traceless perturbation operators by defining the projector in the doubled Hilbert space $\mathbb{P}_0 \!=\! \Dketbra{\mathbb{I}}{\mathbb{I}}/d$ such that $\mathbb{P}_0\Dket{A}\!=\!\Tr(A) \Dket{\mathbb{I}}/d$, and redefine the relevant superoperator 
\begin{equation}
    \tilde{M}_0 = M_0(\mathbb{I}-\mathbb{P}_0).
    \label{eq:redef_superop}
\end{equation}
For any operator $V'$, this acts as
\begin{equation}
\tilde{M}_0\Dket{V'}  = M_0(\mathbb{I}-\mathbb{P}_0)\Dket{V'} = M_0 \Dket{V}= \Dket{\overline{V}_0},
\label{eq:superop_p0}
\end{equation}
where $V$ is a traceless version of $V'$. We remark that any observable conserved under $H_0$ will also be an eigenvector of $M_0$ with  eigenvalue $1$, i.e. it cannot be counteracted due to the limited control terms of the Hamiltonian (see e.g. the discussion in Ref.~\cite{hai2022}). Similar limitations to robustness may apply in the case of other experimental constraints such as pulse intensity and bandwidth limits.

The goal of URC is to minimize the norm of the modified superoperator $\tilde M_0$, which is related to the previous norm as
\begin{equation}
|| \tilde{M}_0||^2 
= || M_0||^2 - \Tr(M_0^\dagger M_0 \mathbb{P}_0) = || M_0||^2 - 1.
\end{equation}

\noindent This allows us to find choices of $U_0$ which yield $\tilde{M}_0\!\simeq\!0$, thus achieving $\Dket{\overline{V}_0}\simeq0$ for any $V$.

To understand how a single solution for $U_0(t)$ can be made robust to arbitrary perturbations, and when this is possible in principle, we note the following connection with unitary designs \cite{gross2007,brown2010,roberts2017}. Discretizing the integral in Eq.~\eqref{eq:V0_def} into $L\!\gg\! 1$ intervals, we find $   \overline{V}_{0} \sim \frac{1}{L} \sum\limits_{k=1}^L U_0^{(k)\dagger} V U_0^{(k)}$, which has the form of an average of the operator $V$ conjugated over a discrete set of unitaries, $U_0^{(k)}$. If the distribution of such unitaries is uniform according to the Haar measure \cite{collins2006}, then the average
\begin{equation}
    \mathbb{E}_{\{U_0^{(k)}\}}[U^\dagger V U] = \frac{1}{d}\Tr(V),
    \label{eq:1design}
\end{equation}
\noindent is known to vanish for all traceless $V$~\cite{collins2006}. A less stringent requirement is for the distribution to only match the first-order moment of the uniform distribution, i.e. to be a 1-design. In fact, since $\mathbb{P}_0\Dket{A}=\Tr(A) \Dket{\mathbb{I}}/d$, we see that the requirement $\tilde{M}_0=0$ immediately implies Eq. (\ref{eq:1design}) for any operator, thus making the path traced by the unitary evolution operator $U_0(t)$ a 1-design. Given that 1-designs exist in SU$(d)$ for any $d$, this connection serves as a formal proof of the existence of URC solutions, i.e. paths in unitary space that achieve perfect target fidelity while being robust to all possible perturbations to leading order~\cite{SupMat}.

Leveraging randomization to increase robustness in quantum processes is routinely done in the context of quantum computing, particularly by tools like dynamical decoupling \cite{zanardi1999,lidar2014review}, dynamically corrected gates \cite{khodjasteh2009,khodjasteh2012,Buterakos2021} and randomized compiling \cite{wallman2016}. Our work shows that, for general quantum systems, it is possible to translate this connection into a requirement on a single object, the superoperator $\tilde{M}_0$, leading to robustness to any perturbation to first order.  As we show in the following, this allows us to set up a quantum optimal control problem to find evolutions that reach a predefined target while at the same time remain robust to arbitrary perturbations.

{\bf \emph{Optimal control.--}}We now demonstrate how URC can be naturally leveraged in numerical optimizations. A generic quantum optimal control (QOC) approach considers a series of control parameters, $\{\phi_k\}$, which determine the time dependence of $H_0(t)$ and aims to maximize the fidelity between a target process $U_{\mathrm{target}}$ and the actual (ideal) evolution operator $U_0(t_f,0)$ by minimizing a cost functional $J_0 \!=\! 1 - F_U(U_{\mathrm{target}},U_0(t_f,0))$ with respect to $\{\phi_k\}$. Additionally, robust QOC usually aims at achieving resilience to perturbations characterized by a known operator $V$. For this task, one can concurrently minimize the fidelity susceptibility given by the control functional $J_V \!=\! \frac{1}{d}||\overline{V}_0||^2$ (see for instance \cite{kosut2022,khodjasteh2012}). Our proposed approach of universally robust QOC instead aims at achieving robustness to an \textit{unknown} error operator $V$. This can be achieved by instead minimizing the functional $J_{\mathrm{U}} = \frac{1}{d}||\tilde{M}_0||^2$~\footnote{In the numerical examples, we use the built-in optimization algorithms of SciPy 1.11 in Python 3. For the single-stage optimization, we resort to the Broyden–Fletcher–Goldfarb–Shanno (L-BGGS-B) algorithm, while for the two-qubit case we implement the Sequential Least Squares Programming (SLSQP) which allows for nonlinear constraints. For the system sizes considered, we found that this numerical computation of the gradient did not significantly affect the optimization time. However, an analytical expression for the gradient of $J_{\text{U}}$ is shown in the Supplementary Material~\cite{SupMat}.}.

\begin{figure}[t!]
    \centering
    \includegraphics[width=1\linewidth]{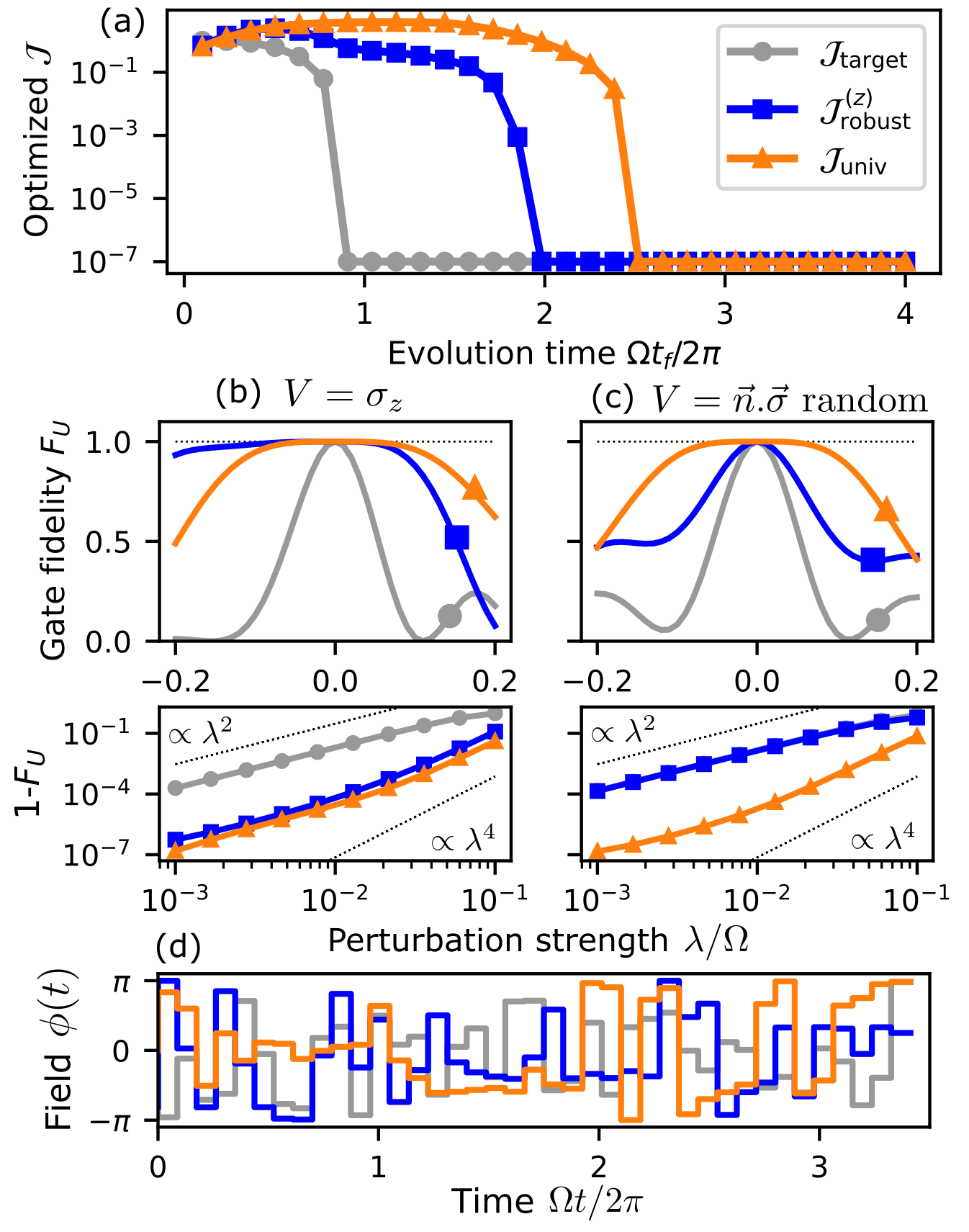}
    \caption{Universally robust control for single-qubit gates. (a) Optimized control functionals as a function of the total evolution time $t_f$ for target-only control (gray, circles), target and robustness to a known $V$ (blue, squares) and target and robustness to an unknown $V$ (orange, triangles). (b) and (c) Gate fidelity as a function of perturbation strength $\lambda$ for the cases where $V=\sigma_z$ and $V=\vec{n} \cdot\vec{\sigma}$ with $\vec{n}$ a random unit vector (results shown correspond to the average fidelity over 20 realizations). Lower panels shows zoomed-in data of the infidelity $1-F$ in log-log scale. (d) Optimal control fields $\phi(t)$ obtained for each case. We choose a target $U_{\mathrm{target}}=\exp(-i \sigma_z \pi/2)$, $N_P=40$ control parameters, a balanced functional $w=1$, and an operation time $\Omega t_f/(2\pi)=3.5$ for (b), (c), and (d).}
    \label{fig:figure1}
\end{figure}

We begin with the simple case of a single qubit with restricted controls with Hamiltonian 
\begin{equation}
    H_0(t) = \Omega\left[\cos [\phi(t)] \sigma_x + \sin [\phi(t)] \sigma_y\right],
    \label{eq:hami_1q}
\end{equation}
where $\sigma_\alpha$ are the Pauli operators, and we consider the control field $\phi(t)$ to be piecewise constant with time steps $\Delta t$ and values $\{\phi_k\}$, $k\!=\!1,\ldots,N_P$ \footnote{The choice of piecewise ansatz is for convenience. Other function bases could equally be used with this method e.g. a Fourier basis, see \cite{SupMat} for numerical results using different parametrizations}. The model in Eq.~\eqref{eq:hami_1q} is fully controllable \cite{boozer2012,poggi2019}.
We set the target transformation to be $U_{\mathrm{target}}\!=\!\exp(-i \sigma_z \pi/2)$ and numerically seek the QOC parameters that minimize either only $\mathcal{J}_{\mathrm{target}}\!=\!J_0$, $\mathcal{J}_{\mathrm{robust}}^{(z)}\!=\!(J_0 + w J_{V=\sigma_z})/(1+w)$ or $\mathcal{J}_{\mathrm{univ}}\!=\!(J_0 + w J_{\mathrm{U}})/(1+w)$, where $w$ is a non-negative weight which can be changed to improve the resulting balance between the terms. Note that evaluating these functionals requires only computing the error-free evolution given by $H_0(t)$, and so no numerical simulations of the perturbed dynamics are required at any stage. In Fig.~\ref{fig:figure1}(a) we plot the optimized functional for each case against the evolution time $t_f$. The curves display behavior reminiscent of Pareto-fronts  \cite{chakrabarti2008,caneva2009}, indicative of the fact that optimization succeeds for sufficiently large $t_f$, but fails when the evolution time becomes too constrained. A minimum control time, $t_{\mathrm{MCT}}$, can be assigned to each process by identifying the minimum value of $t_f$ such that the optimization succeeds (which in this case we take as yielding functional values below $10^{-7}$). For target-only and robust control optimizations, we find $t_{MCT}^{\rm T} \!= \!2\pi/\Omega$ and $t_{MCT}^{\rm R}\!=\! 4\pi/\Omega$ which are consistent with previous analytical and numerical studies \cite{boozer2012,poggi2019}. In contrast, universally robust control demands $t_{MCT}^{\rm U}\!=\!5\pi/\Omega$ (see also \cite{cao2023}). 

To characterize the robustness of these control processes, we study how well the evolution under the perturbed Hamiltonian $H_\lambda(t)=H_0(t)+\lambda V$ is able to achieve the target transformation. Fig.~\ref{fig:figure1} shows the cases for (b) $V=\sigma_z$ and (c) $V=\vec{n}\cdot\vec{\sigma}$ with $\vec{n}$ a randomly chosen unit vector. The gate fidelity is plotted against the uncertainty parameter $\lambda$ for the three types of optimal controls found. All cases yield high fidelities if $\lambda\!=\!0$, but the target-only optimization results (gray) deviate substantially from the ideal value once $\lambda\!\neq\! 0$. In (b), we see that the robust control optimization (blue) is insensitive to perturbations in $V\!=\!\sigma_z$, as expected. But (c) reveals that the same control is sensitive to generic perturbations. Remarkably, the URC solution (orange) is insensitive to perturbations along \textit{any} direction. This holds true even accounting for the faster minimal control times required for the other protocols~\cite{SupMat}. We also highlight that the increase in robustness does not require the use of a more complex control waveform, as can be seen from Fig. \ref{fig:figure1} (d).

{\bf \emph{Generalized robustness.--}}Building upon the superoperator in Eq.~\eqref{eq:redef_superop} we can generalize this framework to optimize for robustness to any desired subset of operators. This is particularly relevant for systems beyond a single qubit where the nature of the noise or inhomogeneity is partially known instead of being completely arbitrary.  Thus, rather than making a control protocol robust to all possible operators $V$, we can instead focus on achieving robustness to a particular set of perturbations, for instance, those generated by local operators. In this case, we are interested in the action of the superoperator, $M_0$, only on this reduced set. The advantage of imposing these generalized robustness requirements is that the optimization is less constrained, as effectively less matrix elements are being minimized. Therefore, it is easier to find good solutions even with restricted control time. For example, the total number of operators for $N$ qubits is $4^N$ while for the set of local operators is only $3 N$.

\begin{figure}[t!]
    \centering
    \includegraphics[width=1\linewidth]{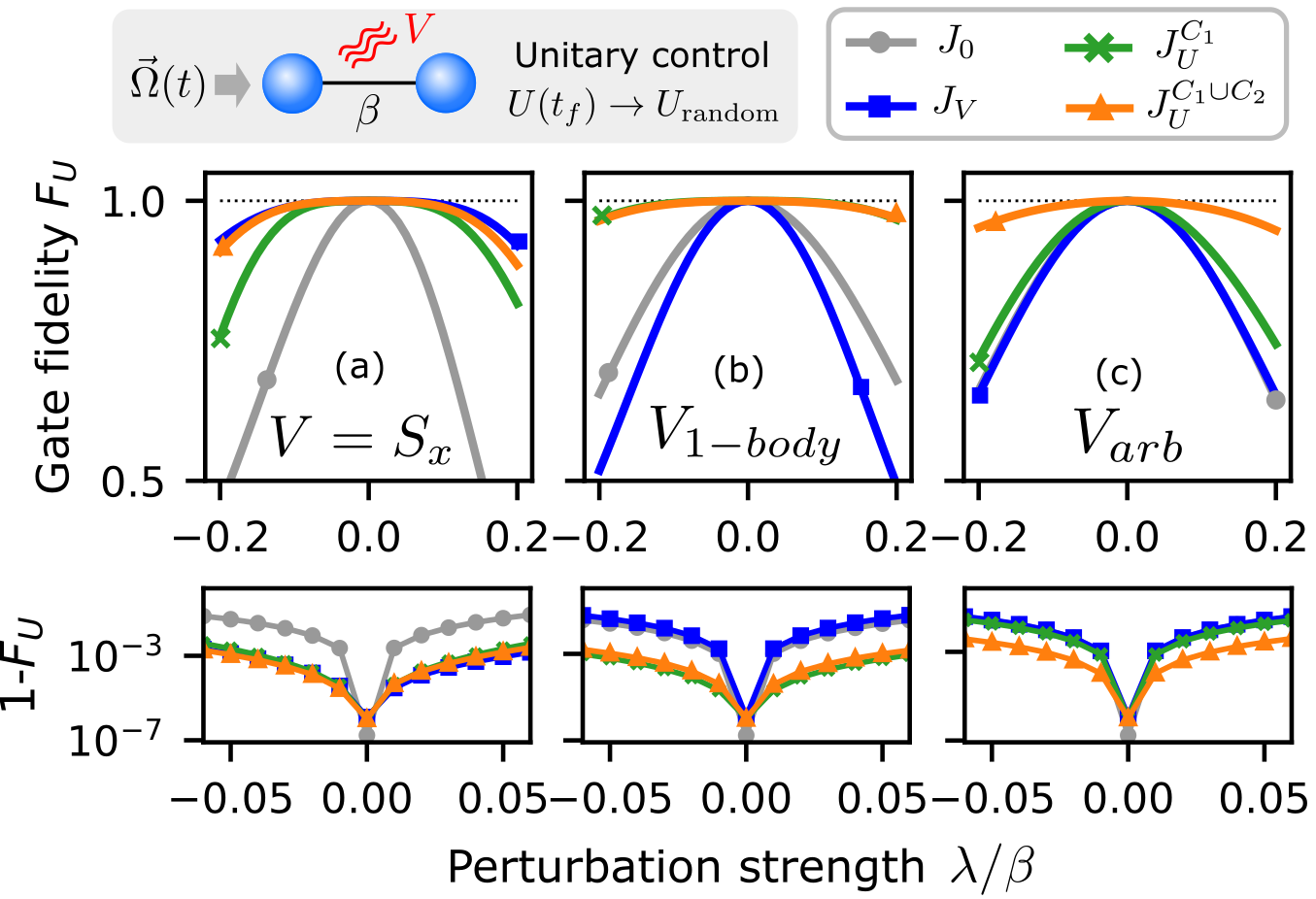}
    \caption{ Universally robust control for two-qubit gates. Plots show the gate fidelity of the perturbed evolution $H_0(t)+\lambda V$, where $H_0(t)$ is the control Hamiltonian of Eq.~\eqref{eq:two_qubit_H0}. Different curves correspond to different types of optimization procedures: target only (nonrobust, gray circles); target and robustness to a fixed $V=S_x$ (blue squares); target and robustness to all single-body operators (green crosses); target and universal robustness (orange triangles). The lower row shows the infidelity $1-F_U$ for each case. Each column shows the performance of each of the four optimized solutions under the evolution $H=H_0(t)+\lambda V$; (a) $V=S_x$, (b) $V$ is random 1-body operator, (c) $V$ is a random arbitrary operator. Evolution time in all cases is $\beta t_f/(2\pi)=5$, and $N_P=50$ control parameters are used. Results shown are averages over $20$ instances.}
    \label{fig:figure2}
\end{figure}

Consider a quantum system with Hilbert space dimension, $d$, and an orthonormal operator basis $\left\{\Lambda_j\right\}$, $j\!=\!0,1,\ldots d^2-1$. We introduce a covering of this basis set, $\{C_k\}$, such that $\left\{\Lambda_j\right\}\!=\!\bigcup_{k=1}^K C_k$. The projector onto $C_k$ is $\mathcal{P}_k(A) \! =\! \sum_{\Lambda_j \in C_k} \Tr(\Lambda_j^\dagger A) \Lambda_j$. In the superoperator picture, this is equivalent to defining $\mathbb{P}_k \!=\! \sum_{\Lambda_j \in C_k} \Dket{\Lambda_j}\Dbra{\Lambda_j}$. These superoperators are clearly projectors, as $\mathbb{P}_k^2 \!=\! \mathbb{P}_k$ and $\sum_{k=0}^K \mathbb{P}_k \!=\! \mathbb{I}$. By construction, we take $\Lambda_0 \!=\! \mathbb{I}/\sqrt{d}$ so that $\mathbb{P}_0$ is defined as before. In order to look for controls which are insensitive to any operator within a given subset we seek to minimize the norm of
\begin{equation}
    \tilde{M}_0 = M_0\left(1-\sum_{k \in \eta}\mathbb{P}_k\right),
    \label{eq:redef_superop2}
\end{equation}
where the sum runs over all relevant operator subsets $\eta$ (typically including $\Lambda_0$). Note that $\mathbb{P}_k$ corresponds to the operators our system's dynamics does not need to be robust to. To illustrate the procedure of imposing generalized robustness requirements into a QOC problem, consider a model of two-qubits with symmetric controls, 
\begin{eqnarray}
    H_0(t)=\Omega_x(t)S_x + \Omega_y(t)S_y + \beta S_z^2,
    \label{eq:two_qubit_H0}
\end{eqnarray}
where $S_{\alpha}\!=\!(\sigma_{\alpha}^{(1)} + \sigma_{\alpha}^{(2)})/2$ are collective spin operators and the interaction strength $\beta>0$ is fixed. The perturbation operator, $V$, can be a combination of single-body ($C_1$) or two-body ($C_2$) operators. We thus have a variety of possible optimization functionals depending on the level of robustness desired. Here we compare three cases: robustness to a single $V\!=\!S_x$, robustness to all single-body operators ($V=V_{1-body}\in C_1$) and universal robustness ($V=V_{arb}\in C_1 \cup C_2$). Here, $V_{1-body}$ and $V_{arb}$ are chosen randomly within the corresponding subspaces. We set the target as a randomly-chosen symmetric two-qubit unitary $U_{\rm random}$ \cite{SupMat}. For this system we find that a good balance between fidelity at zero perturbation and robustness can be achieved by performing a two-stage optimization. First, we minimize the target alone until a certain threshold $J_0< \varepsilon$ is met.  The resulting optimized field is then seeded to the robustness optimization which minimizes $J_V$ or $J_U$ alone, with the added constraint that $J_0$ never exceeds $\varepsilon$.~\footnote{This two-stage approach can be further refined to improve performance, see for instance Ref. \cite{kosut2022}}.
In Fig.~\ref{fig:figure2}, we showcase the performance of the optimization using the different variants introduced thus far, in the presence of various perturbations. As expected, the optimal control procedure is able to find fields which are robust to arbitrary single-body perturbations (green curve), but are not necessarily robust to completely arbitrary perturbations. In contrast, the URC solution (orange curve) results in evolutions which are markedly more robust to any type of perturbation, including two-body operators, when compared to the other methods. 

The approach outlined above for designing generalized robustness requirements can be readily carried over to more complex systems. In the Supplementary Material \cite{SupMat} we show additional results that illustrate how this framework can be used to robustly generate entangled states in many-body systems.

{\bf \emph{Conclusion.--}} We have introduced a versatile method, universally robust control, to mitigate the effects of unknown sources of error. By recasting the impact of an arbitrary perturbation to the systems in terms of a single object, here captured by the superoperator in Eq.~\eqref{eq:bare_suoperop}, we showed that since this superoperator has no explicit dependence on the precise operator form of the error, it can be efficiently minimized to provide the necessary, highly robust, control pulses. This goes beyond previous approaches \cite{green2013,khodjasteh2009,khodjasteh2009pra} since it provides a unifying framework for achieving universal robustness for \textit{arbitrary} finite-dimensional quantum systems, while concurrently defining a concise methodology to implement numerical optimization to achieve robust controls in practice. We demonstrated the effectiveness of our approach for the realization of single- and two-qubit quantum gates, and have shown that it can be generalized to tackle state control problems or to the case of classical fluctuations~\cite{Kiely2021}. Furthermore, we have demonstrated that the URC formalism can exploit partial information about the source of errors to build arbitrary robustness requirements into the optimal control problem. When combined with powerful numerical optimization techniques, we expect this flexible approach  to be able to tackle a broad class of questions in quantum control. 
For instance exploring the fundamental trade-off between robustness and experimental constraints (such as bandwidth or evolution time), or determining what control resources are required to achieve various levels of robustness in a quantum device. Finally, as our protocol introduces control pulses which dynamically implement $1$-designs, this could be generalized to other $t$-designs which can be readily exploited for quantum computing protocols such as randomized benchmarking~\cite{nakata2021}.

\emph{Acknowledgments.--} PMP acknowledges fruitful discussions with Lorenza Viola and Gopikrishnan Muraleedharan. PMP acknowledges support by U.S. National Science Foundation (grant number PHY-2210013) and AFOSR (grant number FA9550-181-1-0064).  GDC acknowledges support by the UK EPSRC EP/S02994X/1. SC acknowledges support from the Alexander von Humboldt Foundation. AK and SC are supported by the Science Foundation Ireland Starting Investigator Research Grant ``SpeedDemon" No. 18/SIRG/5508.\\

\let\oldaddcontentsline\addcontentsline     
\renewcommand{\addcontentsline}[3]{}     

\bibliography{lib}

\pagebreak
\widetext
 
 \newpage 
\begin{center}
\vskip0.5cm
{\Large Supplemental Material}
\end{center}
\vskip0.4cm

\setcounter{section}{0}
\setcounter{equation}{0}
\setcounter{figure}{0}
\setcounter{table}{0}
\setcounter{page}{1}
\renewcommand{\theequation}{S\arabic{equation}}
\renewcommand{\thefigure}{S\arabic{figure}}

\let\addcontentsline\oldaddcontentsline     

\tableofcontents{}
\section{Perturbative approximation to the fidelity \label{A}}

\subsection{Fidelity susceptibility as the quantum Fisher information \label{A1}}

We are interested in calculating the quantum Fisher information (QFI) of the state $\rho_\lambda$ with respect to the parameter $\lambda$. Formally, the QFI is defined as $F_Q[\rho_\lambda]=(\Delta L)^2$,
which is the variance ($(\Delta x)^2=\avg{x^2}-\avg{x}^2$) of the symmetric logarithmic derivative $L$ defined implicitly as $\frac{\partial \rho_\lambda}{\partial \lambda}= \left( \rho_\lambda L + L\rho_\lambda  \right)/2$ \cite{liu2020}. Note that for pure states we have that
\begin{eqnarray}
\frac{\partial \rho_\lambda}{\partial \lambda} &=& \frac{\partial \rho_\lambda^2}{\partial \lambda} \\
&=&   \rho_\lambda \frac{\partial \rho_\lambda}{\partial \lambda} + \frac{\partial \rho_\lambda}{\partial \lambda} \rho_\lambda.
\end{eqnarray}
Clearly then for this case, $L=2 \frac{\partial \rho_\lambda}{\partial \lambda}$. The mean is given by
\begin{eqnarray}
\avg{L} &=& 2 \Tr \left( \rho_\lambda \frac{\partial \rho_\lambda}{\partial \lambda} \right) \\
 &=& \Tr \left(  \frac{\partial \rho_\lambda^2}{\partial \lambda} \right) \\
 &=& \frac{\partial}{\partial \lambda} \Tr \rho_\lambda^2 \\
 &=& 0.
\end{eqnarray}
Therefore the QFI can be compactly written as
\begin{eqnarray}
F_Q[\rho_\lambda]=4 \Tr \left[ \rho_\lambda \left(\frac{\partial \rho_\lambda}{\partial \lambda}\right)^2 \right].
\end{eqnarray}

\subsection{Explicit expression for the fidelity susceptibility \label{A3}}

The derivative of the state can be expressed as
\begin{eqnarray}
\label{eq:app_partialrho}
\frac{\partial \rho_\lambda}{\partial \lambda} = \frac{\partial U_\lambda}{\partial \lambda} \sigma U_\lambda^\dagger + U_\lambda \sigma \frac{\partial U^\dagger_\lambda}{\partial \lambda}.
\end{eqnarray}
We then use the derivative of the unitary time evolution operator as
\begin{eqnarray}
\frac{d U_\lambda(t_f,0)}{d\lambda}= -\frac{i}{\hbar} \int_0^{t_f} ds U_\lambda(t_f,s) V U_\lambda(s,0).
\label{eq:dU_dlambda}
\end{eqnarray}
Inserting in Eq.~\eqref{eq:app_partialrho} gives us
\begin{eqnarray}
\frac{\partial \rho_\lambda}{\partial \lambda} &=& -\frac{i}{\hbar} \int_0^{t_f} ds \left[ U_\lambda(t_f,s) V U_\lambda(s,0) \sigma U_\lambda^\dagger(t_f,0) - U_\lambda(t_f,0) \sigma U^\dagger_\lambda(s,0) V U^\dagger_\lambda(t_f,s) \right] \\
&=& -\frac{i}{\hbar} \int_0^{t_f} ds \left[ U_\lambda(t_f,s) V U^\dagger_\lambda(t_f,s) U_\lambda(t_f,0) \sigma U_\lambda^\dagger(t_f,0) - U_\lambda(t_f,0) \sigma U^\dagger_\lambda(t_f,0) U_\lambda(t_f,s) V U^\dagger_\lambda(t_f,s) \right] \\
&=& -\frac{i}{\hbar} \int_0^{t_f} ds \left[ U_\lambda(t_f,s) V U^\dagger_\lambda(t_f,s) \rho_\lambda - \rho_\lambda U_\lambda(t_f,s) V U^\dagger_\lambda(t_f,s) \right] \\
&=& -i \left[G_\lambda,\rho_\lambda\right],
\end{eqnarray}
where $G_\lambda=\frac{1}{\hbar}\int_0^{t_f}ds U_\lambda(t_f,s) V U^\dagger_\lambda(t_f,s)$. The QFI is then
\begin{eqnarray}
F_Q[\rho_\lambda]&=& - 4 \, \Tr \left\{\rho_\lambda \left[G_\lambda,\rho_\lambda\right]^2 \right\} \\
&=& - 4 \, \Tr \left\{\rho_\lambda \left( G_\lambda \rho_\lambda G_\lambda \rho_\lambda -G_\lambda \rho_\lambda^2 G_\lambda- \rho_\lambda G_\lambda^2 \rho_\lambda+\rho_\lambda G_\lambda \rho_\lambda G_\lambda \right)  \right\} \\
&=& 4 \left( \avg{G^2_\lambda}-\avg{G_\lambda}^2 \right) \\
&=& 4 \left( \Delta G_\lambda \right)^2,
\end{eqnarray}
where these steps hold true for pure states. This can be further simplified by noting that
\begin{eqnarray}
\avg{G_\lambda} &=& \Tr \left\{ G_\lambda \rho_\lambda \right\} \\
&=& \Tr \left\{ \frac{1}{\hbar}\int_0^{t_f}ds U_\lambda(t_f,s) V U^\dagger_\lambda(t_f,s) U_\lambda(t_f,0) \sigma U^\dagger_\lambda(t_f,0) \right\} \\
&=& \Tr \left\{ \frac{1}{\hbar}\int_0^{t_f}ds U^\dagger_\lambda(s,0) V U_\lambda(s,0) \sigma \right\} \\
&=& t_f \avg{\overline{V}_{\lambda}}_i / \hbar,
\end{eqnarray}
where the average $\avg{\cdot}_i$ is taken over the initial state $\sigma$ and we have defined $\overline{V}_{\lambda} = \frac{1}{t_f} \int_0^{t_f} ds U^\dagger_\lambda(s,0) V U_\lambda(s,0)$ which is the time-averaged version of the operator $V$ in the interaction picture. Similarly, the other term is given by
\begin{eqnarray}
\avg{G^2_\lambda} &=& \Tr \left\{ G^2_\lambda \rho_\lambda \right\} \\
&=& \Tr \left\{ G^2_\lambda U_\lambda(t_f,0) \sigma U^\dagger_\lambda(t_f,0) \right\}\\
&=& \Tr \left\{ U^\dagger_\lambda(t_f,0) G^2_\lambda U_\lambda(t_f,0) \sigma  \right\}\\
&=& \Tr \left\{ \left[U^\dagger_\lambda(t_f,0) G_\lambda U_\lambda(t_f,0)\right]^2 \sigma  \right\}\\
&=& t_f^2 \avg{\overline{V}_\lambda^2}_i /\hbar^2.
\end{eqnarray}

Putting this all together, we have that the QFI can be expressed as a variance of the operator $\overline{V}_{\lambda}$ over the initial $\lambda$-independent state $\sigma$,
\begin{eqnarray}
F_Q[\rho_\lambda]&=& \frac{4 t_f^2}{\hbar} \left( \Delta \overline{V}_\lambda \right)^2.
\end{eqnarray}

\subsection{Fidelity susceptibility for unitaries} \label{app:derivation_unitary}

The relevant fidelity is $F_U(\lambda) = \frac{1}{d^2}\left \vert \Tr\left(U_0^\dagger U_\lambda\right)\right\vert^2$ \cite{poggi2019} which we will expand in powers of $\lambda$. Again we make use of Eq. (\ref{eq:dU_dlambda}) and start by noting that for any complex scalar $z$, $\frac{d}{dx}(|z|^2)= 2\mathrm{Re}(z^* \frac{dz}{dx})$. Therefore
\begin{align}
    \frac{dF_U}{d\lambda} &=\frac{2}{d^2}\mathrm{Re}\left[\Tr(U_\lambda^\dagger U_0)\Tr\left(U_0^\dagger \frac{dU_\lambda}{d\lambda}\right) \right]\\
    &= \frac{2}{d^2 \hbar}\mathrm{Im}\left\{ \Tr(U_\lambda^\dagger U_0) \int_0^{t_f} ds\: \Tr\left[U_0^\dagger U_\lambda(t_f,s)V U_\lambda(s,0)\right]  \right\}.
\end{align}

Evaluating the first derivative at $\lambda=0$ yields
\begin{equation}
    \left. \frac{dF_U}{d\lambda} \right\vert_{\lambda=0} = \frac{2}{d \hbar}\mathrm{Im}\left\{ \int_0^{t_f} ds\: \Tr\left[U_0^\dagger(s,0)V U_0(s,0)\right]\right\} = \frac{2}{d \hbar}\mathrm{Im}\left[ \Tr(V) t_f\right] = 0.
\end{equation}

The second derivative reads
\begin{align}
    \frac{d^2 F_U}{d\lambda^2} &= \frac{2}{d^2 \hbar}\mathrm{Im}\left\{
    \Tr\left[\frac{dU_\lambda^\dagger}{d\lambda} U_0\right] \int_0^{t_f} ds\: \Tr\left[U_0^\dagger U_\lambda(t_f,s)V U_\lambda(s,0)\right] \right. \nonumber + \\
    & \left. \Tr(U_\lambda^\dagger U_0) \int_0^{t_f} ds\: \Tr\left[ U_0^\dagger \frac{d U_\lambda(t_f,s)}{d\lambda} V U_\lambda(s,0) + U_0^\dagger U_\lambda(t_f,s)V \frac{d U_\lambda(s,0)}{d\lambda}\right]\right\} \\
    &= \frac{2}{d^2 \hbar^2}\left\vert \int_0^{t_f} ds\: \Tr\left[U_0^\dagger U_\lambda(t_f,s)V U_\lambda(s,0)\right]\right\vert^2  \nonumber  + \\
    & -\frac{2}{d^2 \hbar^2}\mathrm{Im}\left\{ i\Tr(U_\lambda^\dagger U_0) \int_0^{t_f} ds\: \Tr\left[ \int_s^{t_f} dx\: U_0^\dagger U_\lambda(t_f,x) V U_\lambda(x,s) V U_\lambda(s,0) + \int_0^s dx\: U_0^\dagger U_\lambda(t_f,s)V U_\lambda(s,x) V U_\lambda(x,0)\right] \right\}.
\end{align}

When evaluating it at $\lambda=0$, the first term depends on $\Tr(V)$ and, as before, vanishes. For the second term, we use the cyclic property of the trace to find that
\begin{align}
    \left.\frac{d^2 F_U}{d\lambda^2}\right\vert_{\lambda=0} &=-\frac{2}{\hbar^2 d}\mathrm{Im}\left[i\int_0^{t_f} ds\: \left\{ \int_s^{t_f} dx\: \Tr[V(x)V(s)] + \int_0^s dx\: \Tr[V(s)V(x)]\right\}\right] \\
    &= -\frac{2}{\hbar^2 d}\mathrm{Im}\left[i\: \Tr\left( \int_0^{t_f} ds\: V(s) \int_0^{t_f} dx\: V(x)\right)\right]  = -\frac{2 t_f^2}{\hbar^2 d} \Tr\left(\overline{V_0}^2\right).
\end{align}

This is exactly the quoted result in Eq.~\eqref{eq:fide_sus_u}.

\section{Operator Hilbert space}

In this section, we review the concepts and notation used when working in the Hilbert space of operators. An example of this is also discussed in Ref.~\cite{am2015}. This formulation is often used for numerical simulations. In our case, it is employed to clearly represent an important linear map between observables.

The key concept is that the operator $V$ is written as a vector $\Dket{V}$ i.e. the $d \times d$ matrix is now represented by a vector of length $d^2$. The superoperator $\mathcal{M}_{0}$ which maps operators to operators can be now written as a $d^2 \times d^2$ matrix $M_0$. The time averaged $V$ in the interaction picture, which is used to determine the fidelity susceptibility, can be now written as $M_0 \Dket{V}=\Dket{\overline{V_0}}$. The goal is now to find $M_0$ in a particular representation. In our case we use the mapping $\sum\limits_{ij} A_{ij} \ketbra{i}{j} \rightarrow \sum\limits_{ij} A_{ij} \ket{ij}$, with a notation $\ket{ij}=\ket{i}\otimes \ket{j}$.

To show this explicitly, we first consider
\begin{eqnarray}
    U A U^\dagger &=& \sum_{i,j,k,l} U_{i,j} A_{j,l} (U^\dagger)_{l,k} \ketbra{i}{k}\\
    &\rightarrow& \sum_{i,j,k,l} U_{i,j} U^*_{k,l} A_{j,l} \ket{ik} \\
    &=& \sum_{i,j,k,l} U_{i,j} U^*_{k,l} \ketbra{ik}{jl} \sum_{n,m} A_{n,m} \ket{nm} \\
    &=& \left(\sum_{i,j} U_{i,j} \ketbra{i}{j} \right) \otimes \left( \sum_{k,l} U^*_{k,l} \ketbra{k}{l}  \right) \Dket{A} \\
    &=&  U \otimes U^* \Dket{A}.
\end{eqnarray}
Note that if we replace $U$ with $U^\dagger$, we get the result needed i.e. $U^\dagger A U \rightarrow U^\dagger \otimes U^T \Dket{A} = \left[U \otimes U^*\right]^\dagger \Dket{A}$.

In the operator Hilbert space inner products can be written as
\begin{eqnarray}
    \Dbraket{A}{B} &=& \left(\sum_{i,j} A^*_{i,j} \bra{ij} \right) \left( \sum_{n,m} B_{n,m} \ket{nm} \right) \\
    &=& \sum_{i,j,n,m} A^*_{i,j} B_{n,m} \braket{ij}{nm} \\
    &=& \sum_{i,j} A^*_{i,j} B_{i,j} \\
    &=& \sum_{i,j} (A^\dagger)_{j,i} B_{i,j} \\
    &=& \Tr(A^\dagger B).
\end{eqnarray}

The identity matrix in this representation becomes a vector $\Dket{\mathbb{I}}=\sum_{i} \ket{ii}$. The projection in the main text is then $\mathbb{P}_0=\frac{1}{d} \sum_{i,j} \ketbra{ii}{jj}$. Applying this to a general operator gives $\mathbb{P}_0 \Dket{A}=\frac{1}{d}\sum_{i,j,k,l} A_{k,l} \ket{ii}\braket{jj}{kl}=\frac{1}{d}\sum_{i,j} A_{j,j} \ket{ii}= \Tr(A)/d \Dket{\mathbb{I}}$.

\section{Existence of URC solutions and absence of trade-off between fidelity and robustness}

An important aspect of the URC framework is that it analytically shows that there is no fundamental trade-off between robustness and target fidelity (i.e., the fidelity at $ \lambda=0$, absent perturbation). This is because of the existence of solutions (i.e. paths in unitary space) that achieve zero infidelity $J_0=0$ while at the same time being perfectly robust to any perturbation (to quadratic order in perturbation strength), by making the norm of the robustness superoperator $||M_0|| = 1$ (equivalently $|| \tilde{M}_0 || = 0$). 

A formal proof is as follows: $\tilde{M}_0 = 0$ implies that the unitary generated by the control protocol obeys Eq. (\ref{eq:1design}); i.e. that it generates a unitary 1-design (see also the related moment superoperators in, e.g., \cite{brown2010}). Recall, unitary $t$-designs are sets of unitaries which mimic the completely uniform (Haar) distribution up to its $t$-th moments. For the special unitary group SU$(d)$, it is known that unitary 1-designs exist for all $d$ (in fact, unitary $t$-designs in SU$(d)$ exist for all combinations of $t$ and $d$). If the unitary path exists, then there is a Hamiltonian $H(t)$ that generates it.  The main constraint for a 1-design to exist is its number of elements; in our case this is not a problem because the paths generated by the control are continuous and thus can be approximated by a number of points which in general is much larger than $d$. Finally, the control problems of relevance here require the 1-design to have two fixed elements: the initial state $U(0,0)=\mathbb{I}$ and the final target state $U(t_f,0)=U_{\rm target}$. This is also generically achievable for a 1-design. To see why, consider a list $(U_1,U_2,U_3,...,U_M)$ constituting a 1-design. If each element is transformed by a fixed unitary $W$ as $U_i’ = W U_i$, then the modified list $(U_1',\ldots,U_M')$ is also a 1-design, since
\begin{equation}
    \frac{1}{M}\sum_{k=1}^M (U_i')^\dagger V U_i'=\frac{1}{M}\sum\limits_{k=1}^M U_i^\dagger (W^\dagger V W) U_i = \frac{1}{d}\Tr(W^\dagger V W)=\frac{1}{d}\Tr(V) \, \, \forall\: V.
\end{equation}

Then, we can create two new different lists by transforming the original list with unitaries $W_A$ and $W_B$. We choose those unitaries such that $W_A U_1 = \mathbb{I}$ and $W_B U_M = U_{\text{target}}$.
The new lists $(U_i^A)_{i=1}^M$ and $(U_i^B)_{i=1}^M$ with $U_i^A=U_1^\dagger U_i$ and $U_i^B=U_{\text{target}}U_M^\dagger U_i$, are then 1-designs by construction. Finally we construct a list $(\tilde{U}_i)_{i=1}^{2M}$ which includes all the elements of both (in particular $\mathbb{I}$ and $U_{\text{target}}$) where
\begin{eqnarray}
    \tilde{U}_i =\begin{cases}
        U_i^A & i=1,\ldots,M \\
        U_{i-M}^B & i=M+1, \ldots, 2M 
    \end{cases}.
\end{eqnarray}
It is easy to see that this is also a 1-design
\begin{align}
    \frac{1}{2M} \sum_{k=1}^{2M} \tilde{U}_k^\dagger V \tilde{U}_k &= \frac{1}{2M}\left ( \sum_{i=1}^M (U_i^A)^\dagger V U_i^A + \sum_{j=1}^M (U_j^B)^\dagger V U_j^B\right) \\
    &= \frac{1}{2M}\left[\frac{M}{d}\Tr(V) + \frac{M}{d}\Tr(V)\right]=\frac{1}{d}\Tr(V)\ \forall V.
\end{align}

Thus, given two arbitrary elements of SU($d$), it is always possible to construct a 1-design that contains them both. Note that we have explicitly referred to these collections of unitary matrices as lists rather than sets, as there could be repetition present.

This result implies that there is no fundamental trade-off between robustness and target fidelity. In practice, however, several constraints could hinder the degree of control available on a system: limited bandwidth, energy, control time etc. These constraints will introduce a trade-off, necessarily, between target fidelity and robustness. This is because robustness requires more control resources; e.g. as seen from Fig \ref{fig:figure1}, it requires longer control times.  When limited resources are available, a compromise must be met. We consider this to be a universal fact in quantum control problems, and one that is fundamentally unavoidable.

\begin{figure}[h]
    \centering
    \includegraphics[width=0.6\linewidth]{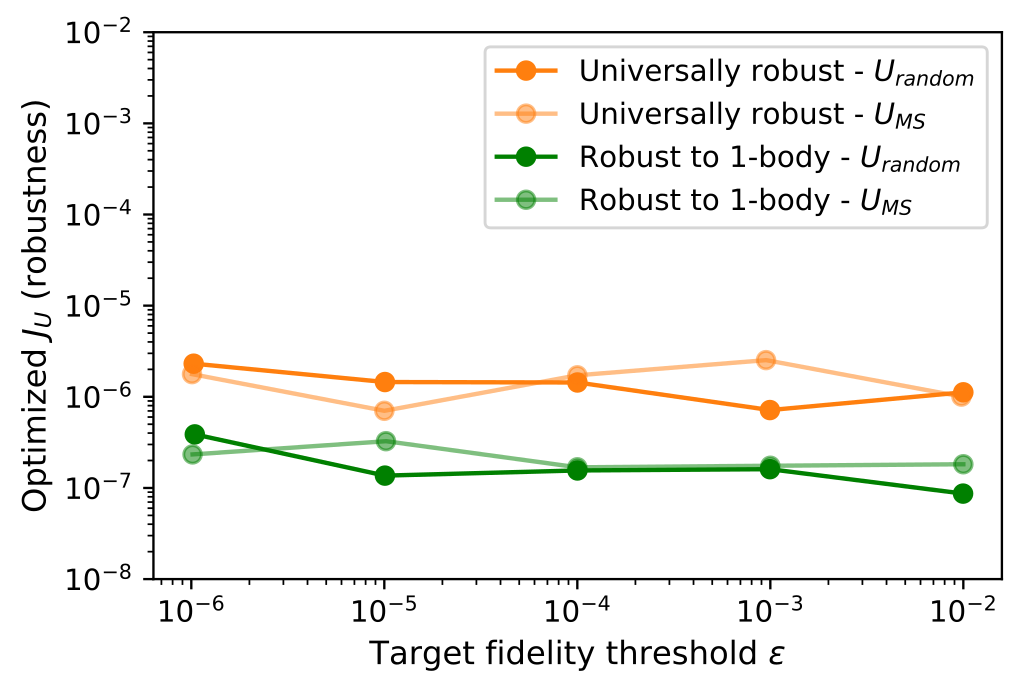}
    \caption{Optimized robustness functional $J_U$ as a function of the target fidelity threshold $\varepsilon$ imposed in the two-stage optimization procedure. Results are shown for two-qubit gate optimization, with two different targets $U_{\rm random}$ and $U_{MS}$ (see Sec. VII.C for more details), and two different levels of robustness: either universally robust or robust to single-body perturbations only. Other parameters are chosen as in Fig. \ref{fig:figure2} of the main text.}
    \label{fig:SM_tradeoff}
\end{figure}

While a systematic analysis of this trade-off between URC and control resources is beyond the scope of this work, we numerically show here that for the cases considered in our work, allowing enough control time is enough to achieve a degree of robustness which is independent of the target fidelity (thus, showing that there is no trade-off between the two). Consider the two-qubit gate optimization treated in the main text. We can explore the performance of the optimization by changing the target fidelity threshold $J_0<\varepsilon$ and studying how well the robustness functional $J_U$ can be optimized. Recall that $J_0<\varepsilon$ also imposes a nonlinear constraint that is built into the second stage of the optimization. We show results of this analysis in Fig.~\ref{fig:SM_tradeoff}. The plot demonstrates that lowering the threshold $\varepsilon$ (i.e. requiring higher and higher nominal fidelities) does not significantly affect the best possible robustness that can be achieved. This is seen both for the complete universal robustness and for an example of generalized robustness requirements, as well for the two types of target gates studied in our work. 

\section{Robust state control}

Coming back to the original state control problem, we now show how the different universal robustness constraints can be imposed in that case too. Recall the expression of the fidelity susceptibility (or quantum Fisher information) associated with the perturbation-dependent trajectories $\rho_\lambda(t)$, Eq. (\ref{eq:state_qfi}). The relevant variance can be written (for a pure initial state) as
\begin{eqnarray}
   \left(\Delta \overline{V}_0\right)^2 
&=& \Tr[\overline{V}_0^2  \sigma]-\Tr[\overline{V}_0 \sigma \overline{V}_0 \sigma] \\
&=& \Dbra{\overline{V}_0 }\mathbb{I}\otimes \sigma^* \Dket{\overline{V}_0} - \Dbra{\overline{V}_0} \sigma \otimes \sigma^* \Dket{\overline{V_0}}\\
   &=& \Dbra{V} M_0^\dagger \mathbb{P}_\sigma M_0 \Dket{V},
\end{eqnarray}
where we have defined $\mathbb{P}_\sigma=(\mathbb{I}-\sigma) \otimes \sigma^*$. This is clearly a projector since 
\begin{eqnarray}
    \mathbb{P}_\sigma^2 &=& (\mathbb{I}-\sigma)^2 \otimes (\sigma^*)^2 \\
    &=& (\mathbb{I}-\sigma-\sigma+\sigma) \otimes \sigma^*  \\
    &=& \mathbb{P}_\sigma.
\end{eqnarray}

We will now simplify this projector further. Its eigenvectors $\Dket{\chi_n}$ with eigenvalue one, must satisfy $\mathbb{P}_\sigma \Dket{\chi_n} \rightarrow (\mathbb{I}-\sigma) \chi_n \sigma=\chi_n$. For pure states this simplifies to
\begin{eqnarray}
     \left[ \chi_n -\Tr(\sigma \chi_n)\right]\sigma= \chi_n.
\end{eqnarray}
Multiplying on the right by $\sigma$ we can see that the condition reduces to $\Dbraket{\sigma}{\chi}=0$. The projection operator can be written as  
\begin{eqnarray}
   \mathbb{P}_\sigma = \sum_n \Dketbra{\chi_n}{\chi_n},
\end{eqnarray}
where the operators $\chi_n$ form an orthogonal basis and which all fulfil $\Dbraket{\sigma}{\chi_n}=0$. Written simpler $\mathbb{P}_\sigma=\mathbb{I}-\Dketbra{\sigma}{\sigma}$, the latter expression leads to
\begin{eqnarray}
    \left(\Delta \overline{V}_0\right)^2 = \norm{\overline{V}_0}^2-\left|\Dbraket{\sigma}{\overline{V}_0}\right|^2.
\end{eqnarray}
Note also that by the Cauchy–Schwarz inequality $    \left(\Delta \overline{V}_0\right)^2 = \norm{\mathbb{P}_\sigma M_0 \Dket{V}}^2 \leq \norm{M_0 \Dket{V}}^2 = \norm{\overline{V}_0}^2$. This connects to the case of quantum gates since the last term is state independent, c.f. Eq. \eqref{eq:fide_sus_u}.

To summarise then, we have
\begin{equation}
\left(\Delta \overline{V}_0\right)^2 = \Dbra{V}(M_0^\sigma)^\dagger M_0^\sigma  \Dket{V
}.
\label{eq:varV_superop}
\end{equation}
which has the same form as Eq. \eqref{eq:normV_superop}, but introduces the initial-state-dependent superoperator $M_0^\sigma  = \mathbb{P}_\sigma M_0 $. Note that this does not suffer the same difficulty as before since $M^\sigma_0\Dket{\mathbb{I}}=\mathbb{P}_\sigma M_0 \Dket{\mathbb{I}}=\mathbb{P}_\sigma \Dket{\mathbb{I}}=0$.

\section{Analytical expression of the URC functional gradient}

In the following, we will derive a closed form approximation to the gradient of the URC cost functional $J_{\mathrm{U}}$, which could be then used in a variety of gradient based optimization algorithms. 

\subsection{The gradient of a unitary evolution operator}

Let us assume the ideal Hamiltonian to be split as a drift and a time-dependent control part: $H(t)=H_{\rm d}+H_{\rm con}(t)$.  The time evolution operator under this Hamiltonian is formally $U(t,0)= \mathcal{T} \exp\left[-\frac{i}{\hbar} \int_0^t H(s) ds\right]$, where $\mathcal{T}$ is the time ordering operator. We assume that the time dependence is piecewise constant, i.e.,  on the interval $[t_j ,t_{j+1}]$ we have $H_{\rm con}(t)=\phi_j W$.  The time evolution operator can then be expressed exactly as
\begin{eqnarray}
U(t,0) &=& \prod_j \exp\left[-\frac{i}{\hbar} (H_{\rm d} + \phi_j W) \Delta t \right] \\
&=&  \prod_j U_j ,
\end{eqnarray}
where $U_j$ is the time evolution operator over constant intervals $\Delta t = t_{j+1}-t_j$. We can define the control vector as $\vec{\phi}=(\phi_1, \phi_2,  \ldots, \phi_{N_P})^T$.  The $k^{\rm th}$ element of the gradient of the time evolution operator with respect to this control vector can be expressed as
\begin{eqnarray}
\{\nabla_{\vec{\phi}} U(t,0) \}_k
&= & \prod_{j=k+1}^{N_P} U_j \left(\partial_{\phi_k} U_k \right)    \prod_{j=1}^{k-1} U_j \\
&\approx & \prod_{j=k+1}^{N_P} U_j \left( - \frac{i}{\hbar} \Delta t W U_k \right)    \prod_{j=1}^{k-1} U_j,
\end{eqnarray}
where in the second step we have assumed short time intervals $\Delta t$.
The gradient over any interval $[t_a,t_b]$ on our mesh can be expressed as
\begin{eqnarray}
    \{\nabla_{\vec{\phi}} U(t_b,t_a) \}_k \approx \begin{cases}
    0 & t_k \notin [t_a,t_b] \\
    \prod_{j=k+1}^{b} U_j \left( - \frac{i}{\hbar} \Delta t W U_k \right)    \prod_{j=a}^{k-1} U_j & \rm{Otherwise}
\end{cases} \label{gradU}.
\end{eqnarray}

\subsection{The gradient of the cost functional $J_{\mathrm{U}}$}
We now want to find the gradient of the norm (we assume the Frobenius norm for concreteness) of the superoperator.  The norm squared can be first simplified as
\begin{eqnarray}
d J_{\mathrm{U}}
&=& ||M_0||^2 -1  \\
&=&   \norm{\frac{1}{t_f} \int\limits_0^{t_f} ds U(s,0) \otimes U(s,0)^*  }^2 -1\\
&=& \frac{1}{ t_f^2}  \Tr \left\{  \int\limits_0^{t_f} ds_2 U(s_2,0) \otimes U(s_2,0)^* \int\limits_0^{t_f} ds_1 \left[U(s_1,0) \otimes U(s_1,0)^*\right]^\dagger  \right\}-1 \\
&=& \frac{1}{t_f^2} \int\limits_0^{t_f} ds_1 \int\limits_0^{t_f} ds_2   \Tr \left[  U(s_2,0) U^\dagger(s_1,0) \otimes U(s_2,0)^*  U(s_1,0)^T  \right]-1 \\
&=& \frac{1}{t_f^2} \int\limits_0^{t_f} ds_1 \int\limits_0^{t_f} ds_2   \Tr \left[  U(s_2,s_1) \otimes U(s_2,s_1)^*  \right] -1\\
&=& \frac{1}{ t_f^2} \int\limits_0^{t_f} ds_1 \int\limits_0^{t_f} ds_2    \Tr \left[  U(s_2,s_1)  \right] \Tr \left[  U(s_2,s_1)^*  \right]-1 \\
&=& \frac{1}{ t_f^2} \int\limits_0^{t_f} ds_1 \int\limits_0^{t_f} ds_2  \left| \Tr \left[  U(s_2,s_1)  \right] \right|^2 -1.
\end{eqnarray}
Note that by the Cauchy-Schwartz inequality and the fact that the time evolution operator is unitary we get that $J_{\rm U} \leq d-1/d$.  

The $k^{\rm th}$ component of the gradient is then
\begin{eqnarray}
\partial_{\phi_k} J_{\mathrm{U}} &=& \frac{1}{d t_f^2} \int\limits_0^{t_f} ds_1 \int\limits_0^{t_f} ds_2  \partial_{\phi_k} \left| \Tr \left[  U(s_2,s_1)  \right] \right|^2 \\
&=& \frac{2}{d t_f^2} \int\limits_0^{t_f} ds_1 \int\limits_0^{t_f} ds_2  \mathrm{Re}\left\{\Tr \left[  U(s_2,s_1)  \right]  \partial_{\phi_k} \Tr \left[  U(s_2,s_1)  \right]^* \right\} \\
&=& \frac{2}{d t_f^2} \int\limits_0^{t_f} ds_1 \int\limits_0^{t_f} ds_2  \mathrm{Re}\left\{\Tr \left[  U(s_2,s_1)  \right]  \Tr \left[ \partial_{\phi_k} U(s_2,s_1)  \right]^* \right\}.
\end{eqnarray}

This could be computed numerically using the following steps. First compute all $U_j$ for a given vector $\vec{\phi}$. Then, the derivative can be approximated as
\begin{eqnarray}
\partial_{\phi_k} J_{\mathrm{U}} \approx \frac{2 (\Delta t)^2 }{d t_f^2} \sum_{n,m=1}^{N_P-1} \mathrm{Re}\left\{\Tr \left[  U(t_m,t_n)  \right]  \Tr \left[ \partial_{\phi_k} U(t_m,t_n)  \right]^* \right\} ,
\end{eqnarray}
where $U(t_m,t_n)= U_m \ldots U_n$ and each component of the gradient of $U$ is approximated by Eq. \eqref{gradU}.

In order to reduce the computation time in calculating $U_j$, one could use the Baker-Campbell-Hausdorff approximation as
\begin{eqnarray}
    U_j \approx \exp\left[- \frac{i}{\hbar}(\Delta t) H_{\rm d}\right] \exp\left[- \frac{i}{\hbar}(\Delta t) \phi_j W\right] \exp\left[\frac{1}{2\hbar^2}(\Delta t)^2 \phi_j [H_{\rm d}, W]\right],
\end{eqnarray}
provided that $\Delta t$ was sufficiently small. Precomputing the spectrum of $H_{\rm d}$, $W$ and $[H_{\rm d},W]$ would make the repeated matrix exponentiation much faster. 

\section{Extension to classical fluctuations}

Consider now the Hamiltonian $H(t)=H_0(t)+\lambda \xi(t) V$. The noise averaged state fidelity is given to second order in $\lambda$ as
\begin{eqnarray}
    \avg{F_\xi} \approx 1-\frac{\lambda^2}{\hbar^2} \int_0^{t_f} dt \int_0^{t_f} ds C(t,s) \left[ \avg{V_I(t)V_I(s)}-\avg{V_I(t)}\avg{V_I(s)}\right],
\end{eqnarray}
where $V_I(t)=U_0^\dagger(t,0)V U_0(t,0)$ is the noise operator $V$ in the interaction picture and the noise $\xi(t)$ has zero mean and correlation function $C(t,s)=\avg{\xi(t)\xi(s)}$. We can define a superoperator $N_t=\left[U_0(t,0) \otimes U^*_0(t,0) \right]^\dagger$, such that $\Dket{V_I(t)}=N_t \Dket{V}$. The first term in the operator Hilbert space can then be expressed as
\begin{eqnarray}
    \avg{V_I(t)V_I(s)}=\Dbra{V}N_t^\dagger \mathbb{I} \otimes \sigma^*  N_s \Dket{V}.
\end{eqnarray}
Similarly the product of averages becomes
\begin{eqnarray}
    \avg{V_I(t)}\avg{V_I(s)} &=&  \Tr[V_I(t) \sigma V_I(s) \sigma]\\
&=& \Dbra{V} N_t^\dagger \sigma \otimes \sigma^* N_s \Dket{V}.
\end{eqnarray}
All together then, this can be written as
\begin{eqnarray}
    \avg{F_\xi} \approx 1-\frac{\lambda^2}{\hbar^2} \int_0^{t_f} dt \int_0^{t_f} ds \, C(t,s) \Dbra{V} N_t^\dagger \mathbb{P}_\sigma N_s \Dket{V}.
\end{eqnarray}
Thus, to minimise the impact of the noise regardless of the operator $V$, one must minimise the operator
\begin{eqnarray}
    \int_0^{t_f} dt \int_0^{t_f} ds \, C(t,s)  N_t^\dagger \mathbb{P}_\sigma N_s.
\end{eqnarray}
This is related to the concept of filter functions \cite{green2013}. Note however that our results are applicable to arbitrary Hamiltonians.

\section{Additional numerical results}
In this section, we present additional numerical results. These include further applications of the URC framework and more detailed descriptions of the problems analyzed in the main text.

\subsection{Generation of many-body entangled states}

\begin{figure}[h]
    \centering
    \includegraphics[width=0.6\linewidth]{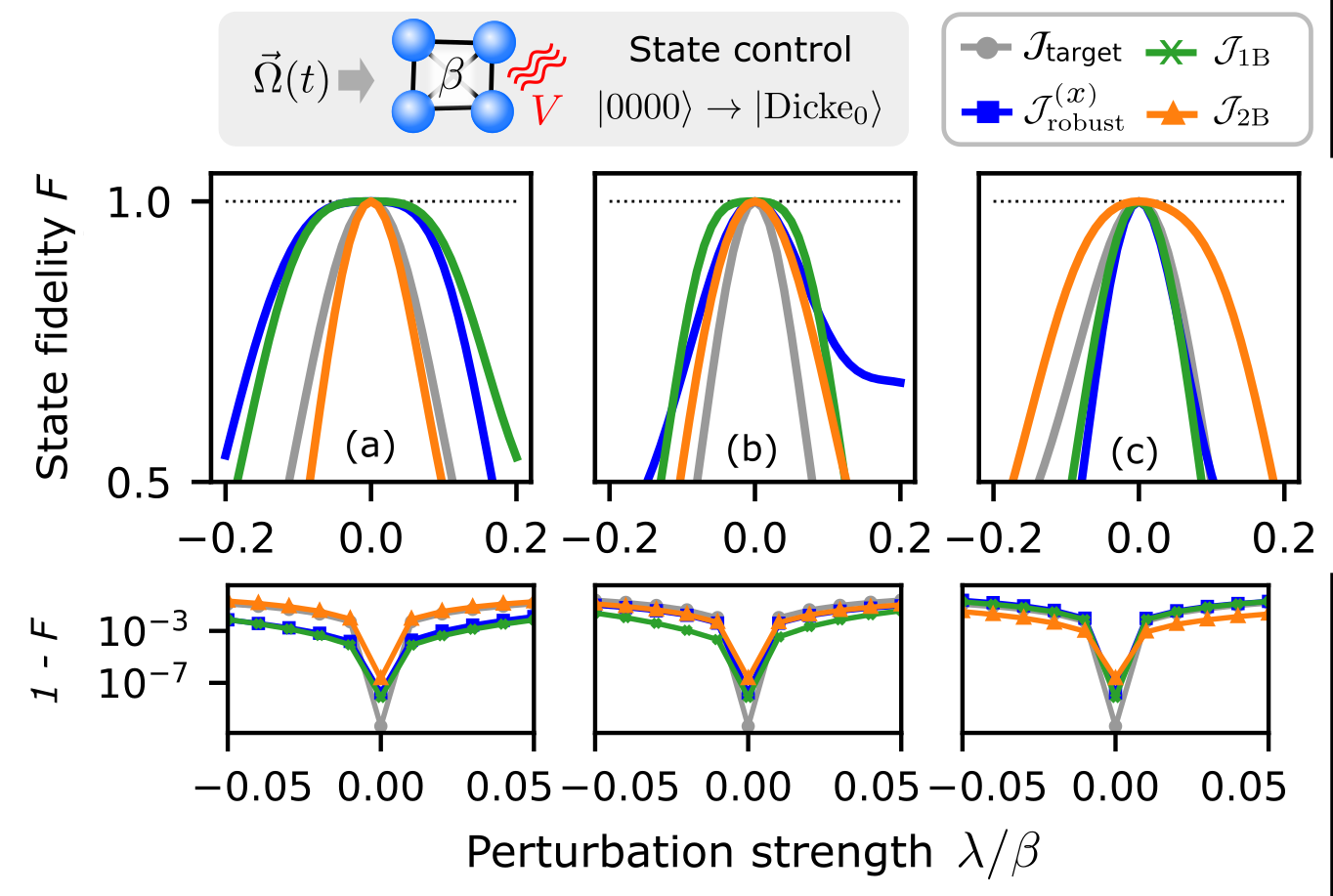}
    \caption{ Universally robust control for the four-qubit state-control problem. Top panels show the state fidelity $F$, bottom panels show a zoomed-in view of the infidelity $1-F$. In all cases data is plotted against the perturbation strength $\lambda$ for the evolution $H_0(t)+\lambda V$ where (a) $V=S_x$, (b) $V=S_z$, (c) $V=S_x^2$. The four curves correspond to the four optimization functionals described in the text, c.f. Eqs.~(\ref{eq:SM_func1})-(\ref{eq:SM_func3}).}
    \label{fig:SM_statecontrol}
\end{figure}

Eq. (\ref{eq:varV_superop}) allows us to carry over the optimal control procedure discussed in the main text for unitary control to the problem of robust state control. The only adaptation needed is to replace $M_0$ with $M_0^\sigma$. We illustrate this procedure by analyzing the problem of generating entangled states in a system of $N=4$ qubits with global controls and all-to-all interactions. We consider a Hamiltonian having the exact same form as (\ref{eq:two_qubit_H0}) where now $S=N/2$ is the total angular momentum associated with the symmetric subspace of the $N$ particles. We point out that this problem is fully controllable for any $N$ \cite{merkel2008}. We consider the problem of driving the system from the state $\ket{0000}$ to the Dicke-0 state, i.e. the eigenstate of $S_z$ composed of a symmetric superposition of states with equal number of qubits in 0 and in 1. Because this is a four-body system, there are many possible choices of robustness setups that could be pursued. Here we demonstrate the flexibility of our approach by showing results corresponding to robustness to \textit{either} all single-body or all two-body operators in Fig. \ref{fig:SM_statecontrol}. In all plots we show four curves, corresponding to four functionals being optimized. These are
\begin{align}
    \mathcal{J}_{\mathrm{target}}&=J_0 \label{eq:SM_func1}\\ \mathcal{J}_{\mathrm{robust}}^{(x)} &= (J_0 + w J_{V=S_x})/(1+w),\label{eq:SM_func2} \\ \mathcal{J}_{\eta} &= (J_0 + w J_{\mathrm{U}}^{(\eta)})/(1+w),\ \text{where}\ \eta = \text{1B},\ \text{2B}. \label{eq:SM_func3}
\end{align}
The URC functionals $J_{\mathrm{U}}^{(\eta)}=||\tilde{M_0}^{(\eta)}||^2/d$ differ in the case of seeking robustness to just 1-body operators (1B) or just 2-body operators (2B):
\begin{align}
   \tilde{M_0}^{\text{(1B)}} &= M_0\left(1-\sum\limits_{k\neq 1} \mathbb{P}_k\right),\\ 
   \tilde{M_0}^{\text{(2B)}} &= M_0\left(1-\sum\limits_{k\neq 2} \mathbb{P}_k\right),
\end{align}

\noindent where $\mathbb{P}_k$ denotes the projector onto $k-$body operators. Fig. \ref{fig:SM_statecontrol} shows the fidelity as a function of perturbation strength $\lambda$ for every solution in the presence of three different perturbations; in other words we calculate the state fidelity achieved by the dynamics of $H_0(t)+\lambda V$ where the different panels correspond to (a) $V=S_x$, (b) $V=S_z$, (c) $V=S_x^2$ (these are the same choices used for the two-qubit case of Fig. \ref{fig:figure2}). The results show that every optimization delivers the expected results: the usual robust control is only insensitive to the predefined choice of $V$, but remains sensitive to other perturbations. On the other hand, the URC waveforms designed to be insensitive to all single-body perturbations (1B) shows robustness in both cases (a) and (b). Likewise, the URC waveforms for $\eta=\text{2B}$ are only robust to the case where the noise is on a two-body operator as in (c). 

\subsection{Comparison of timescales}

\begin{figure}[h]
    \centering
    \includegraphics[width=0.55\linewidth]{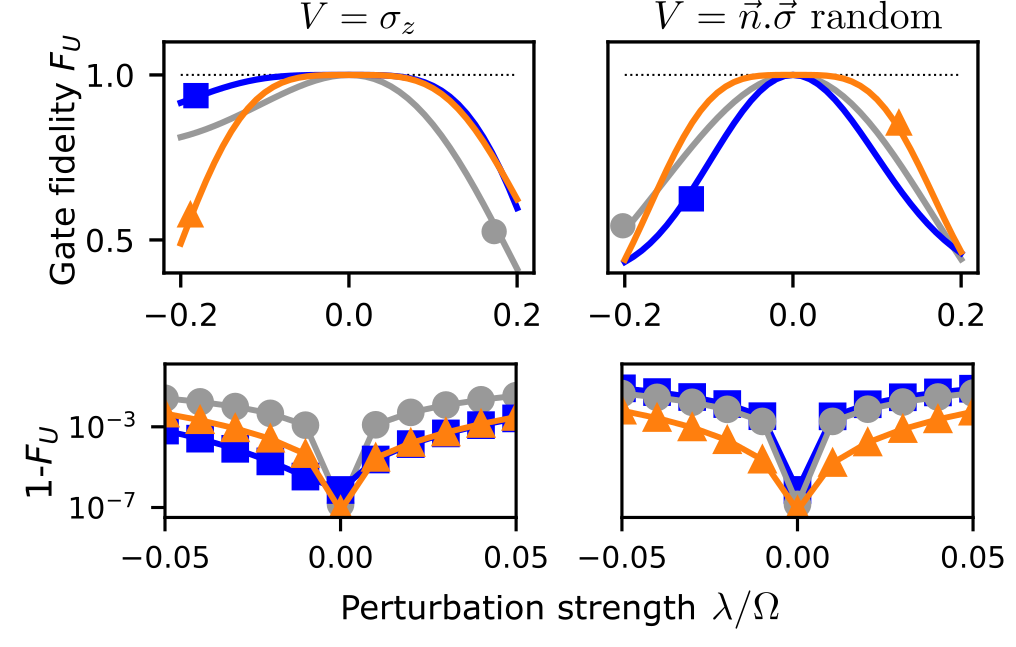}
    \caption{ Performance of universally robust control when compared to other approaches and restricting the evolution time. The parameters are as in Fig. \ref{fig:figure1}, with the exception that for URC (orange triangles) $\Omega t_f/(2\pi)=3.5$; for the standard robust control $\Omega t_f/(2\pi)=2.1$, for the nonrobust control $\Omega t_f/(2\pi)=1.1$}.
    \label{fig:supmat_1qubit}
\end{figure}

In Fig. \ref{fig:figure1}, we demonstrated how the URC waveforms leads to enhanced robustness with respect to perturbations when compared to the other controls analyzed. These correspond to the output of optimizing either $\mathcal{J}_{\text{target}}$ or $J_{\text{robust}}$. The evolutions studied in Fig. \ref{fig:figure1} correspond to all waveforms of the same duration $\Omega t_f/(2\pi)=3.5\pi$. A fair critique of this analysis is that nonrobust or standard robust waveforms can be achieved with shorter operation times. Therefore we compare in Fig. \ref{fig:supmat_1qubit} the performance of waveforms of different durations, which now scale with their respective minimal control time (indicated in the main text). While less striking, it is nevertheless clear from these results that the URC method still provides additional stability overall, despite needing extra time to do so.

\subsection{Other target states}

\begin{figure}[h]
    \centering
    \includegraphics[width=0.55\linewidth]{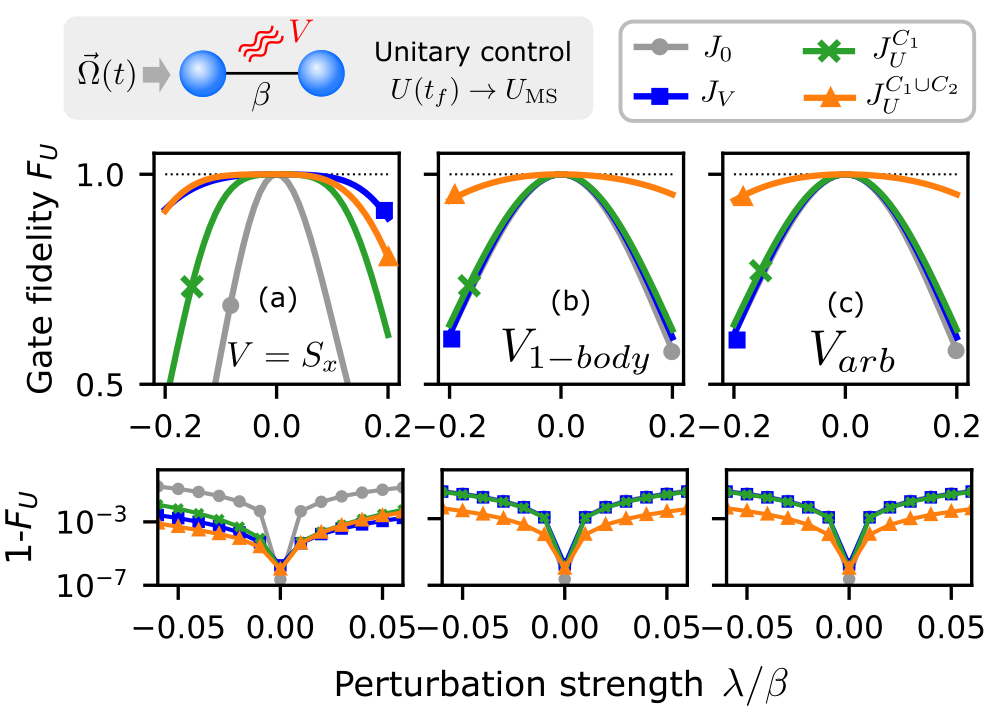}
    \caption{ Universal robust control for two-qubit gates. Plots show the gate fidelity of the perturbed evolution $H_0(t)+\lambda V$, where $H_0(t)$ is the control Hamiltonian of Eq.~\eqref{eq:two_qubit_H0}. Different curves correspond to different types of optimization procedures: target only (nonrobust, gray circles); target and robustness to a fixed $V=S_x$ (blue squares); target and robustness to all single-body operators (green crosses); target and universal robustness (orange triangles). Lower row shows the infidelity $1-F$ for each case.}
    \label{fig:supmat_MSgate}
\end{figure}

In the analysis of two-qubit unitary control, we set as a target transformation a single, randomly-chosen, two-qubit symmetric gate. When written in the symmetric basis $\{\ket{00},(\ket{01}+\ket{10})/\sqrt{2},\ket{11}\}$, such gate has the form
\begin{equation}
    U_{\text{random}} = \left( 
    \begin{array}{c c c}
    0.51762131+0.11456864i & -0.5988566 -0.16086483i &         -0.57589678+0.05271048i \\
    -0.22709248+0.22335233i & 0.30541094+0.57529237i &        -0.6568961 -0.20686492i \\
    -0.75950102+0.20160146i & -0.40091574-0.17470746i &        -0.13888378+0.41469292i
    \end{array}\right).
    \label{eq:app_random_target}
\end{equation}

A more physically-inspired choice could be the XX M\o lmer-S\o rensen (MS) gate,
\begin{equation}
    U_{\text{MS}} = \exp\left[-i\frac{\pi}{2}\left(S_x^2 -\frac{1}{2}S_x\right)\right].
\end{equation}

In Fig. \ref{fig:supmat_MSgate} we show results which are completely analogous to Fig. \ref{fig:figure2}, but now setting as a target the MS gate. As is evident from the data, the URC framework works irrespectively of the choice of target gate. 

\subsection{Optimized control waveforms}
In this section, we show the control waveforms obtained from the numerical optimization which leads to the gate fidelities shown in the main figures. We then show further results with different parametrizations of the control fields, which lead to analogous performances in terms of fidelity and robustness. This showcases how the URC framework can be applied directly to various optimal control approaches.

\begin{figure}[t]
    \centering
    \includegraphics[width=0.8\linewidth]{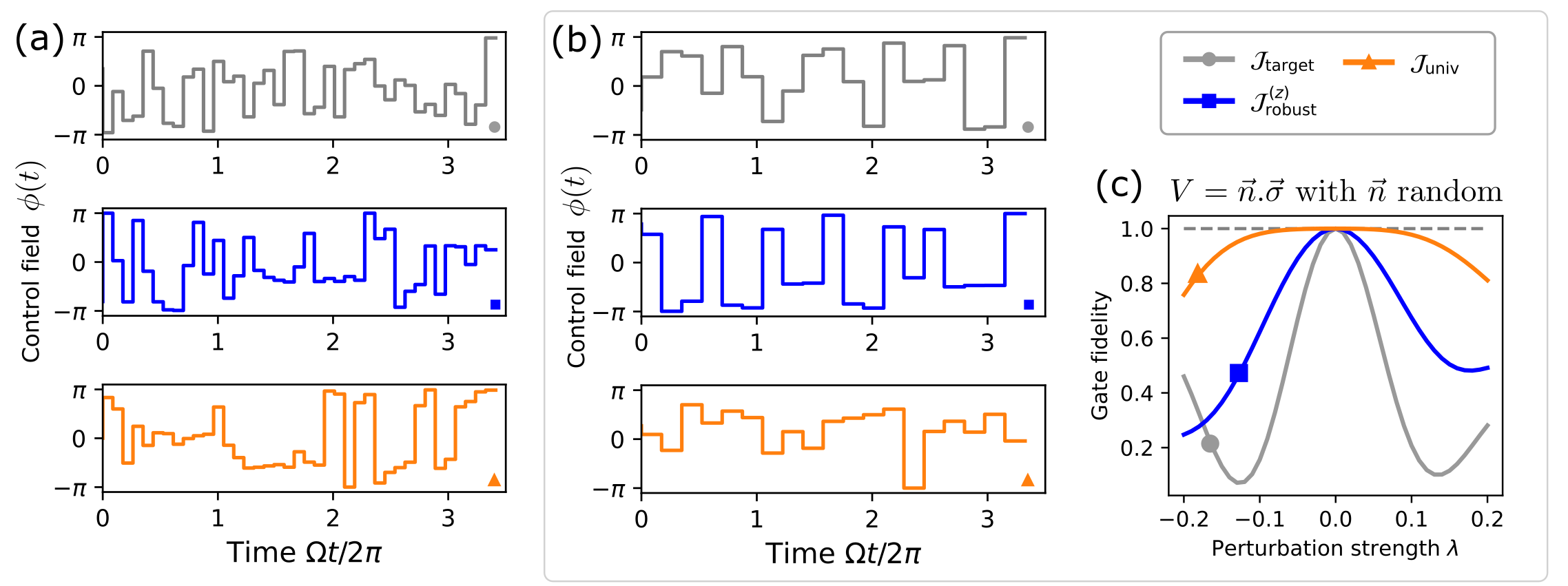}
    \caption{Optimized control waveforms for single qubit gate control and different level of robustness. In all cases, gray (circles) corresponds to nonrobust waveforms, blue (squares) to waveforms robust to perturbation only along Z axis $(V=\sigma_z)$, and orange (triangles) to universally robust waveforms. Panel (a) shows control fields used to construct Fig. \ref{fig:figure1} (b) and \ref{fig:figure1} (c) in the main text, which used $N_p=40$ time steps. Panel (b) shows waveforms obtained by further constraining $N_p = 20$, which perform similarly in terms of fidelity. This can be seen in panel (c) where we plot the average fidelity attained by each of the three waveforms with $N_p = 20$ in the presence of a perturbation with random direction, as a function of the perturbation strength.}
    \label{fig:supmat_waveform_qubit}
\end{figure}

We begin by analyzing the single qubit case. Fig.~\ref{fig:supmat_waveform_qubit} shows various instances of optimal control fields $\phi(t)$ that achieve the target gate described in the main text with various levels of robustness. Panel (a) shows the fields that were used to construct Fig.~\ref{fig:figure1}. These were found by parametrizing the field as a piecewise constant function with $N_p=40$ intervals. The optimal parameters were found by minimizing each of the cost functionals described the main text, and correspond to nonrobust (gray), robust to a given $V=\sigma_z$ (blue) and universally robust (orange). We note two important aspects of these waveforms. First, all of them look similar in that no apparent structure is visible in them. Crucially, there is no change in waveform complexity as the robustness increases; in fact, the universally robust waveform is virtually indistinguishable from the other optimal fields. Second, the minimization with $N_p=40$ is certainly not unique. In fact, we show in (b) a case where we use half of the parameters, $N_p=20$, and thus obtain simpler waveforms. These turn out to have very similar performance in terms of fidelity, as can be seen from Fig. \ref{fig:supmat_waveform_qubit} (c), which can be compared with Fig. \ref{fig:figure1} (c).

\begin{figure}[h]
    \centering
    \includegraphics[width=0.95\linewidth]{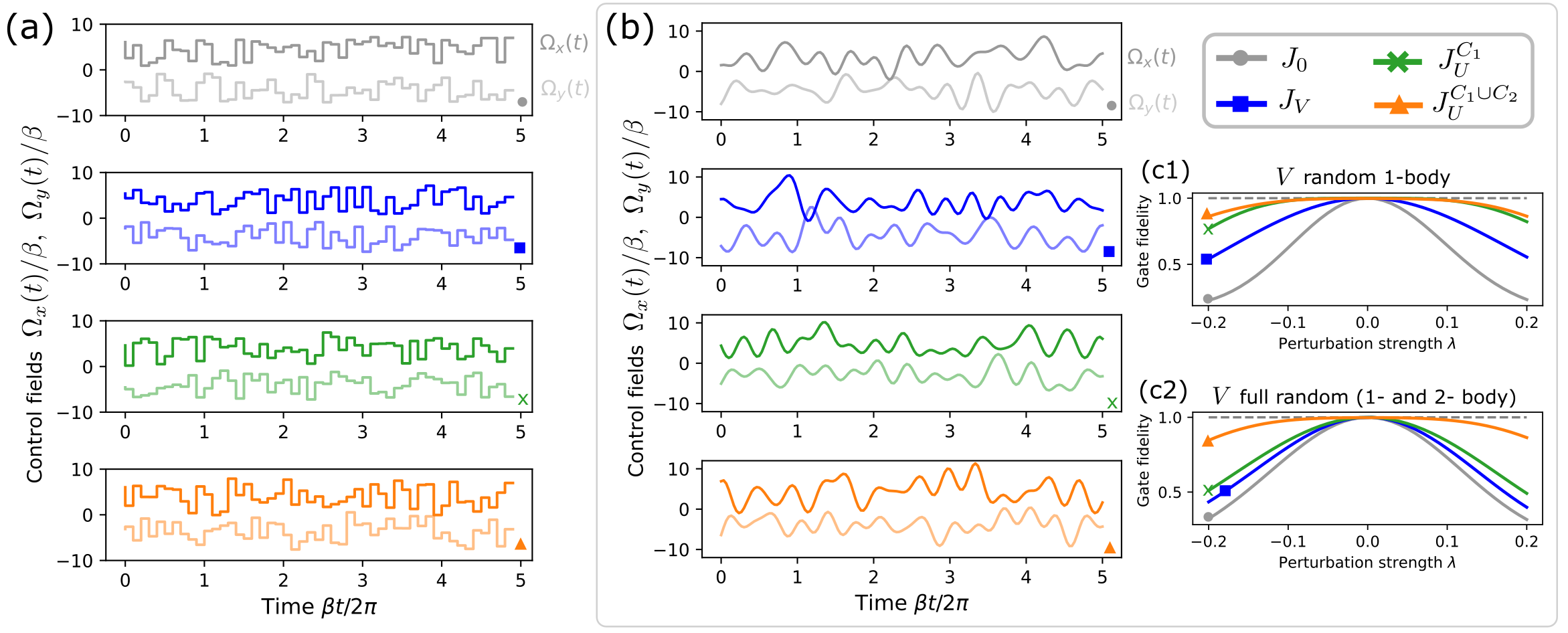}
    \caption{Optimized control waveforms for two-qubit gate control and different level of robustness. In all cases gray (circles) corresponds to nonrobust waveforms, blue (squares) to waveforms robust to perturbation only along X axis $(V=S_x)$, and orange (triangles) to universally robust waveforms. This model involves two control fields $\Omega_x(t)$ and $\Omega_y(t)$; in the figures we plot $\Omega_x(t)+4\beta$ (dark lines) and $\Omega_y(t)-4\beta$ (light lines) so that each field is clearly visible.
    Panel (a) shows control fields used to construct Fig. \ref{fig:figure2} in the main text, which used $N_p=50$ time steps. Panel (b) shows waveforms obtained by changing to a Fourier parametrization. These perform similarly in terms of fidelity, as can seen in panels (c1) and (c2) where we plot, as function of the perturbation strength, the average fidelity attained by each of the four optimized Fourier waveforms in the presence of two types of perturbation $V$: random single-body operator (c1) and fully random (single- and two-body) operator in (c2).}
    \label{fig:supmat_waveform_twoqubits}
\end{figure}

We now turn to the two-qubit gate case. Fig. \ref{fig:supmat_waveform_twoqubits} (a) shows the optimal waveforms that were used to produce the analysis in Fig. \ref{fig:figure2}. These correspond to optimization to achieve a random symmetric two-qubit gate (see Eq. \ref{eq:app_random_target}) which used $N_p=50$ and a total time of $\beta t_f = 5\times (2\pi)$. Note that this model has two independent fields $\Omega_x(t)$ and $\Omega_y(t)$ which are optimized independently. The waveforms shown in Fig. \ref{fig:supmat_waveform_twoqubits} (a), similarly to what we observed for single qubits, are all virtually indistinguishable, and thus again we observe that the robustness requirement does not lead to an increase in waveform complexity. While piecewise constant waveforms can be seen as ``complicated'' from a practical point of view, we stress that this is highly platform dependent. In fact, the waveforms shown here have the same level of complexity as those shown in, e.g. Refs. \cite{smith2013,anderson2015}, which are works demonstrating the experimental implementation of optimal control fields. 

Nevertheless, it is a valid concern that for some experimental platforms it would be more convenient to use other types of waveforms. We stress that the entirety of our URC framework is completely independent of the details of the optimization procedure and of the model specifications, and can thus be applied to approaches beyond piecewise constant fields. We demonstrate this by tackling the robust two-qubit gate problem with an entirely different parametrization, where each of the control fields are now built from a Fourier decomposition,

\begin{equation}
    \sum\limits_{n=0}^{N_{\rm comp}-1} A_n\sin(\omega_n t) + B_n\cos(\omega_n t)
\end{equation}
where we choose to fix the frequencies as $\omega_n = \frac{n}{N_{\rm comp}-1} \omega_{\rm max}$, with $n=0,1,\ldots, N_{\rm comp}-1$. Thus, this parametrization leads to $N_p=2 N_{\rm comp}-1$ variables for each field. We explore this choice of parametrization for the two-qubit robust control model using $N_{\rm comp}=15$ and $\omega_{\rm max}=3\beta$. We run optimizations with the same methods used in the main text (e.g., same target gate, optimization routine, and cost functions). An example of optimal control waveforms found for each case is depicted in Fig.~\ref{fig:supmat_waveform_twoqubits}~(b). We observe that we can find smooth waveforms (by construction) which perform comparably to the piecewise constant ones. This can be seen from the plots in Figs. \ref{fig:supmat_waveform_twoqubits} (c1) and (c2), which depict the average fidelity as a function of perturbation strength obtained with the optimized Fourier fields, for the cases of (c1) a perturbation which is a random 1-body operator, and (c2) a fully random perturbation. Finally, we stress that there is no significant change in the complexity of the waveform that guarantees universal robustness as compared to nonrobust ones.

\subsection{Further details on the numerical optimization}
Finally we comment on the choice of the weight parameter $w$ in the optimization functional for the single-stage optimization procedure, i.e.,

\begin{equation}
    \mathcal{J}_{\mathrm{univ}}\!=\!(J_0 + w J_{\mathrm{U}})/(1+w).
    \label{eq:param_w}
\end{equation}

\begin{figure}[h!]
    \centering
    \includegraphics[width=0.8\linewidth]{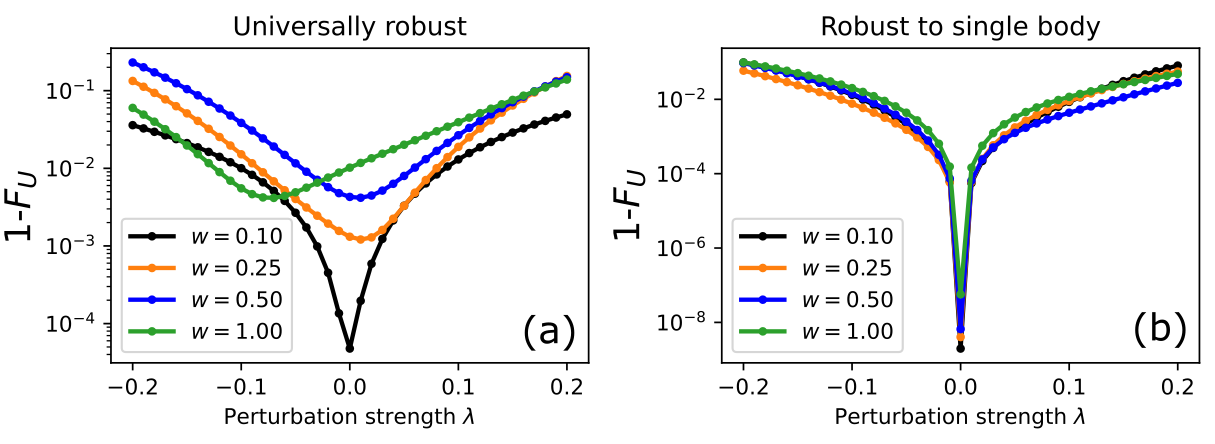}
    \caption{ Analysis of the weight $w$ between functionals in the numerical optimization, see Eq. (\ref{eq:param_w}). Gate infidelity $1-F_U$ as a function of the perturbation strength. Results are shown for optimal control of two qubit gates (same case as shown in the main text) for the cases of (a) Universally robust and (b) Robust to single-body perturbations. Here $V=S_z$ and $\beta t_f = 5$. See text for further details.}
    \label{fig:supmat_param_w}
\end{figure}

For any $w$, the global minimum is $\mathcal{J}_{\mathrm{univ}}=0$ which happens if and only if $J_0 = 0$ and $J_{\mathrm{U}}=0$. When any other solution is found, we would like that both targets are equally well achieved. We find that the straightforward choice of $w=1$ can put too much weight on the robustness requirement, in such a way that the ideal target fidelity found by the optimizer can be too low. This is seen in Fig. \ref{fig:supmat_param_w} (a), where we show the infidelity for the case of two-qubit gates (and random target unitary) when searching for gates that are universally robust. From the data, it can be seen that the infidelity at zero perturbation $\lambda=0$ can be quite high for relatively high values of $w$. This is because the optimizer is not able to get $J_U$ down to values which are of the order of the required infidelities. We expect that this can be improved by allowing more control time, but we leave a more detailed analysis for future work. For a fixed control time, this behavior can be adjusted by lowering the value of $w$. We point out that easier cases, for example the two-qubit case with robustness to only single-body operator, typically do not have a strong dependence on the value of $w$, see Fig. \ref{fig:supmat_param_w} (b).

\end{document}